\documentclass[runningheads]{llncs}

\usepackage{amsmath}

\usepackage{amssymb} 
\usepackage{graphicx}
\usepackage{comment}
\usepackage{subfig}
\usepackage{soul}
\usepackage{enumitem}
\usepackage{color}
\usepackage{multirow}
\usepackage{adjustbox}
\usepackage{xcolor}
\usepackage{listings}

\usepackage{pifont}
\newcommand{\cmark}{\ding{51}}%
\usepackage{url}
\usepackage{fancybox}

\newcommand{\hljs}[1]{{\sethlcolor{orange}\hl{#1}}}

\definecolor{mGreen}{rgb}{0,0.6,0}
\definecolor{mGray}{rgb}{0.5,0.5,0.5}
\definecolor{mPurple}{rgb}{0.58,0,0.82}
\definecolor{backgroundColour}{rgb}{0.95,0.95,0.92}

\lstdefinestyle{CStyle}{
    backgroundcolor=\color{backgroundColour},   
    commentstyle=\color{mGreen},
    keywordstyle=\color{magenta},
    numberstyle=\tiny\color{mGray},
    stringstyle=\color{mPurple},
    basicstyle=\footnotesize,
    breakatwhitespace=false,         
    breaklines=true,                 
    captionpos=b,                    
    keepspaces=true,                 
    numbers=left,                    
    numbersep=5pt,                  
    showspaces=false,                
    showstringspaces=false,
    showtabs=false,                  
    tabsize=2,
    belowskip=3em,
    language=C
}

\setlist[enumerate]{leftmargin=3mm}
\setlist[itemize]{leftmargin=3mm}

\usepackage{mdwlist}

\makecompactlist{itemize}{stditemize}

\usepackage{caption} 
\begin{document}

\title{Attack of the Clones: Measuring the Maintainability, Originality and Security of Bitcoin `Forks' in the Wild}
\titlerunning{Attack of the Clones}
\author{Jusop Choi\inst{1} \and Wonseok Choi\inst{1} \and William Aiken\inst{1} \and Hyoungshick Kim\inst{1} \and Jun Ho Huh\inst{2} \and Taesoo Kim\inst{3} \and Yongdae Kim\inst{4} \and Ross Anderson\inst{5}}
\authorrunning{Jusop Choi et al.}
%
\institute{Sungkyunkwan University, Republic of Korea \and Samsung Research, Republic of Korea \and Georgia Institute of Technology, USA \and Korea Advanced Institute of Science and Technology, Republic of Korea \and Cambridge University, UK}

\maketitle



\begin{abstract}
  Since Bitcoin appeared in 2009, over 6,000 different cryptocurrency projects have followed. The cryptocurrency world may be the only technology where a massive number of competitors offer similar services yet claim unique benefits, including scalability, fast transactions, and security. But are these projects really offering unique features and significant enhancements over their competitors? To answer this question, we conducted a large-scale empirical analysis of code maintenance activities, originality and security across 592 crypto projects. We found that about half of these projects have not been updated for the last six months; over two years, about three-quarters of them disappeared, or were reported as scams or inactive. We also investigated whether 11 security vulnerabilities patched in Bitcoin were also patched in other projects. We found that about 80\% of 510 C-language-based cryptocurrency projects have at least one unpatched vulnerability, and the mean time taken to fix the vulnerability is 237.8 days. Among those 510 altcoins, we found that at least 157 altcoins are likely to have been forked from Bitcoin, about a third of them containing only slight changes from the Bitcoin version from which they were forked. As case studies, we did a deep dive into 20 altcoins (e.g., Litecoin, FujiCoin, and Feathercoin) similar to the version of Bitcoin used for the fork. About half of them did not make any technically meaningful change -- failing to comply with the promises (e.g., about using Proof of Stake) made in their whitepapers.
\end{abstract}



\section{Introduction}
\label{sec:intro}



As of 2021, over 6,000 cryptocurrencies are available for trading on the market~\cite{coinmarketcap} and have attracted attention from securities regulators. However, many cryptocurrency projects (hereinafter referred to as {\em crypto projects}) stopped their development~\cite{rocco,whittemore} and rarely introduce any code updates~\cite{zetzsche2017ico}. A 2018 article in the Wall Street Jurnal claimed that 16\% of 3,291 initial coin offering (ICO) projects show signs of plagiarism, identity theft and promises of improbable goals~\cite{wallstreet}. Worse yet, according to Deadcoins~\cite{deadcoins} and Coinopsy~\cite{coinopsy}, about 1,000 different crypto projects have failed to meet their goals~\cite{scamcoin}. There seems to be little motivation for companies to implement the promised features after their ICO, or to fix security vulnerabilities quickly, because it is not easy for regular investors to validate the claimed features or announced updates by examining the code commits in their project repositories.






In this paper, we claim that (1) many commercially traded crypto projects are not being appropriately maintained, leave known vulnerabilities unpatched, or do not implement promised features, and (2) to improve market transparency and reliability, it is necessary to provide an automated method and tool to regularly assess projects' originality, liveness and security. 

We developed a framework to collect crypto projects' code and project information and analyze the code maintenance liveness, originality, and security. Our framework and dataset will help facilitate more efficient and precise analyses of emerging crypto projects. Our contributions are summarized below:

\begin{itemize}
    \item We collected 592 crypto projects and analyzed their code maintenance data. We found that 288 crypto projects (48.6\%) have not been updated at all for the last six months, and about three-quarters of these disappeared, or were reported as scams or inactive at the end of two years (see Section~\ref{sec:results_maintenance}). 
    \item We conducted a study of security patches in Bitcoin derivatives with over 1.6 million code commits. We developed a code analysis tool to identify security vulnerabilities in projects' Git repositories, and the times at which they were patched. For 11 vulnerabilities patched in Bitcoin, over 80\% of the projects we studied still leave at least one unpatched. The {\em mean time to patch} is 237.8 days (see Section~\ref{sec:results_vulnerability}). 
    \item We evaluated forked projects' originality compared to their parent project. Our results show that over a third of the 157 altcoins that are likely to have been forked from Bitcoin still have 90\% or more code similarity with the Bitcoin version used to spawn them. We did a deep dive into 20 altcoins that are very similar to a Bitcoin version, discovering that about half of them do not contain any technically meaningful update and fail to comply with technical promises -- including about tokenization and Proof of Stakes (PoS) -- made through their whitepapers (see Section~\ref{sec:results_originality}).
\end{itemize}


\section{Evaluation framework}
\label{sec:methodology}


We present a framework for (1) measuring the frequency of crypto project maintenance; (2) identifying security vulnerabilities, and investigating how quickly they are patched or not; and (3) measuring code similarity between crypto projects. Our analysis results are presented in Sections~\ref{sec:results_maintenance},~\ref{sec:results_vulnerability}, and~\ref{sec:results_originality}, respectively.



\subsection{Code maintenance activities analysis}
\label{subsec:maintenance_method}

During the lifetime of any software project, maintenance is required to fix bugs and add new functionality. Crypto projects are no different. A number of metrics can be used to measure maintenance effort. 


We specifically used 32 features obtained from GitHub (for more details, see Appendix~\ref{sec: Features used for evaluating maintenance efforts}). These features can be categorized into three groups: (1) metrics for developers' engagement, (2) metrics for popularity, and (3) metrics for code updates. For the developers' engagement level, we use Commits, Branches, Releases, Contributors, Pull Requests, and Mean Developer Engagement (MDE)~\cite{adams2008detecting}. For project popularity, we analyze Watch, Star, Fork, Issues, Open Issues, and Closed Issues. Finally, for code update status, we use additions, deletions, and time intervals between commits. We use the mean and standard deviation values of these features over 3, 6, and 12 months to evaluate the code maintenance activities in the short-term, mid-term, and long-term, respectively. Although these metrics do not necessarily reflect software quality, Coleman et al.~\cite{Coleman94:software} demonstrated that quantitative metrics such as lines of code can be useful in evaluating software maintainability. Similarly, a recent study~\cite{Coelho20:Software} showed that such code commit-related attributes can be used to evaluate the maintenance level of open-source projects.






\subsection{Security vulnerability analysis}
\label{subsec:vulernability_method}

In crypto projects, it is important to fix security vulnerabilities because of the direct risk of financial loss for users and investors. Therefore, we set out to identify which security vulnerabilities exist and how soon they are patched.

We examine source code for known vulnerabilities and note when each vulnerability is patched. It would be inefficient to download all previous versions of an altcoin's source code from its Git repository and check for vulnerable and patched code. We found that the total size of all previous versions of the crypto projects we analyzed is over 10.2 TB. We therefore developed a technique to inspect only source code changes available on GitHub for each code commit rather than looking at the full source -- we check whether deleted code contains a vulnerability or added code contains a patch. The process of our implementation is summarized in Appendix~\ref{appendix:Process of security vulnerability analysis}.

\subsection{Code similarity analysis}
\label{subsec:originality_method}

To detect and deter plagiarism -- heavy reuse of original code without adding new features -- we compare source code between projects to measure their similarity. If the source code of project $P_1$ has high similarity with that of project $P_2$, and $P_1$ was released after $P_2$, we assess whether a significant portion of $P_1$'s code has been reused without modification from $P_2$.


As most crypto projects are (hopefully) updated over time, it might not be accurate to compare two crypto projects at one random point in time to assess the originality of an altcoin against its parent project. Two projects could appear very different if the parent project underwent significant maintenance, even if the child had not. For example, the source code of Carboncoin, which was forked from Bitcoin on March 8th 2014, would be vastly different from the Bitcoin project snapshot taken on September 4th 2019 (33.5\%), even though Carboncoin has only been slightly updated. This is because Bitcoin has undergone significant maintenance over that five-year period. Therefore, the forked project's originality must be measured against a snapshot of the original parent project taken at the time of forking. We developed two heuristic methods to estimate the fork date of a given project based on the commits in Git. 

\textbf{Heuristic 1:} If a project $P_\textnormal{forked}$ is forked from its original project $P_\textnormal{original}$ in Git, they would have the same code commit history before that forking event. However, their commit history is likely to be different from the time of the fork onward. Based on this intuition, we first determine whether a given crypto project $P_\textnormal{forked}$ is forked from another project $P_\textnormal{original}$, by checking whether the first few commits of the project $P_\textnormal{forked}$ are the same as the commits of the project $P_\textnormal{original}$. We then look for the first different commit history between the two projects and use that time as a forking point. 

\textbf{Heuristic 2:} We observed that some crypto projects were created by externally uploading source code files of another project. For those projects, we cannot use the first heuristic because we cannot compare common commits. In such cases, however, we can often observe the largest number of files being added or updated at a specific time to upload source code files of the parent project. Based on this intuition, in a given crypto project, $P_\textnormal{forked}$, we first identify which commit contains the largest number of file additions and/or changes from the project's commit. We use this commit time as a hypothetical (likely) forking point for project $P_\textnormal{forked}$ from its original project $P_\textnormal{original}$. We then sequentially compute the similarity between the source code of $P_\textnormal{forked}$ at the time of the fork and previous versions of source code of $P_\textnormal{original}$, respectively, before the time of the fork to find the most similar previous version of $P_\textnormal{original}$. This is because a crypto project can often be forked from an out-of-date version of its parent project. In principle, we need to examine all previous versions of $P_\textnormal{original}$ (before the time of the fork to find the source code version of $P_\textnormal{original}$) maximizing the similarity between $P_\textnormal{forked}$ and $P_\textnormal{original}$. However, to reduce computation cost, we only considered all previous versions for the last 6 months from the hypothetical (likely) forking point. Finally, if the computed similarity of the commit having the highest similarity score is greater than a predefined threshold, we assess that project $P_\textnormal{forked}$ was forked from project $P_\textnormal{original}$ with high probability. 






For a code-similarity scoring tool, we used \texttt{JPlag}~\cite{jplag,prechelt2002finding} because it has become known for its robustness against many of the techniques used to disguise similarities between plagiarized files~\cite{ko2017coat}. 



\subsection{Dataset}
\label{subsec:dataset}


For evaluation, we initially collected all 1,627 crypto projects registered on CoinMarketCap~\cite{coinmarketcap} as of July 2018. However, because we were only interested in cryptocurrency projects that operate independently on their own blockchains, we did not include token projects (e.g., ERC-20, ERC-223, and ERC-721 types) in our analysis. Finally, we selected 592 crypto projects\footnote{The list is online at \url{https://bit.ly/3A0ppjl}.} that were registered as an independent blockchain platform, and for which the official GitHub URLs were available. We started from the list of registered cryptocurrencies on July 22nd, 2018 but downloaded the latest version before September 3rd, 2019 of their source code files. Since our goal was to study the long-term (1--2 years) project development and maintenance activities and trends of cryptocurrency projects, we selected projects that were created around 2018 and analyzed their historical data for the following 1--2 years. Further, we needed ground-truth information about the survivability of the projects (i.e., whether projects were eventually reported as scams, or inactive, or were still operating) to investigate the relationship between projects' maintenance activities and some form of ground truth about project survivability.

Crypto projects are implemented in a wide variety of programming languages including C, Go, Python, and Java. However, we found that 511 (86.3\%) of the 592 crypto projects we obtained from our selection process are written in C/C++. Because the archetypal crypto project -- Bitcoin -- is written in C/C++, we focused on C-based crypto projects for our vulnerability and code similarity analyses. For the code maintenance analysis, we used all 592 projects because maintenance may not be highly related to the underlying programming language.



\section{Code maintenance activity analysis}
\label{sec:results_maintenance}

This section presents the results of code maintenance activity analysis, which was performed using the method described in Section~\ref{subsec:maintenance_method}. We used the $k$-means clustering algorithm to categorize 592 crypto projects into $k$ groups by code maintenance activity to detect poorly maintained or inactive code development. In our dataset, the best clustering results were obtained when $k=4$ (see Appendix~\ref{appendix:Selection of k in k-means clustering}). The clustering results are summarized in Table~\ref{table:Clustering results of cryptocurrencies by code maintenance activities}.

\begin{table}[!ht]
\vspace{-0.7cm}
    \centering
    \caption{Clustering results of cryptocurrencies by code maintenance activity during 2018. Examples are selected by market capitalization.}
    \resizebox{0.74\linewidth}{!}{
        \begin{tabular}{|c|c|c|c|}
            \hline
            ID & \# & Examples & Properties \\ \hline
            1 & 207 & DGB, LSK, XZC, LBC, GAP & rarely updated during the last 12 months\\ \hline
            2 & 81 & RDD, FST, TTC, MOON, SUMO & rarely updated during the last 6 months\\ \hline
            3 & 61 & XEM, SC, PZM, UNO, SLS & rarely updated during the last 3 months\\ \hline
            4 & 243 & BTC, ETH, XRP, LTC, ADA & frequently updated during all periods\\ \hline
        \end{tabular}}
    \label{table:Clustering results of cryptocurrencies by code maintenance activities}
\vspace{-0.5cm}
\end{table}

\begin{figure}[!ht]
    \vspace{-0.2cm}
    \centering
    \begin{minipage}[b]{0.31\textwidth}
        \centering 
        \includegraphics[width=\textwidth]{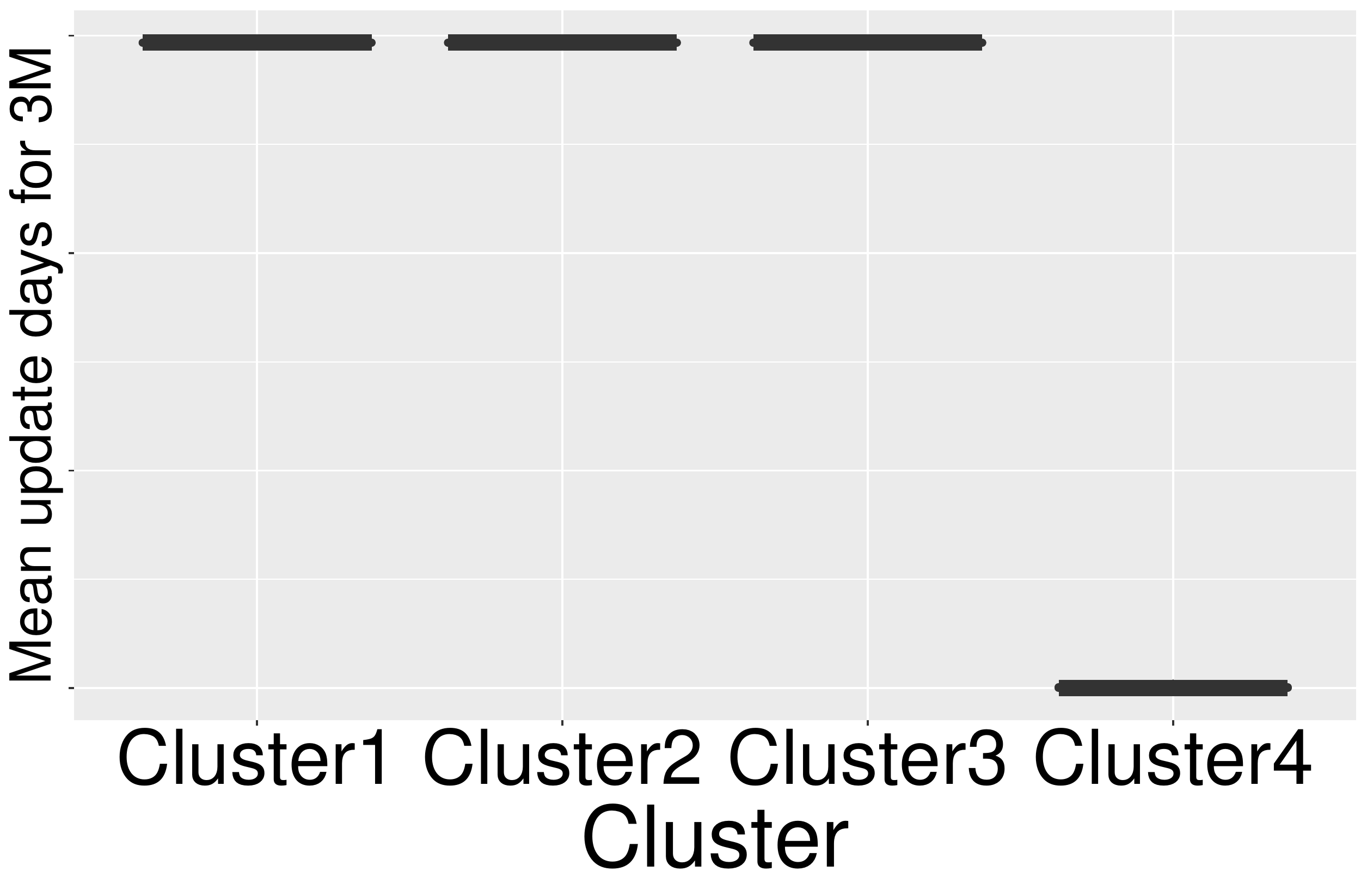}
    \end{minipage}\hfill
    \begin{minipage}[b]{0.31\textwidth}
        \centering 
        \includegraphics[width=\textwidth]{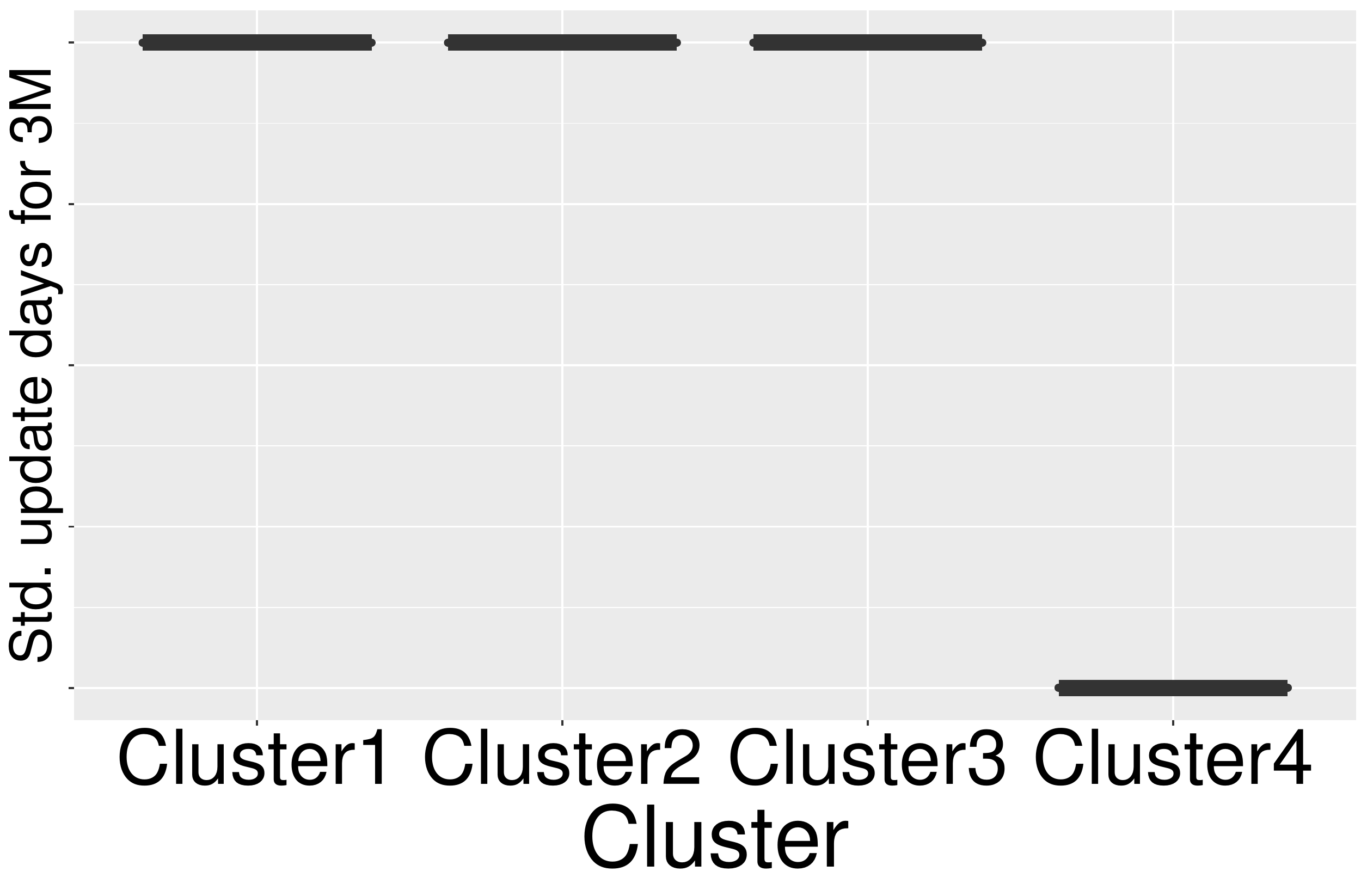}
    \end{minipage}\hfill
    \begin{minipage}[b]{0.31\textwidth}
        \centering 
        \includegraphics[width=\textwidth]{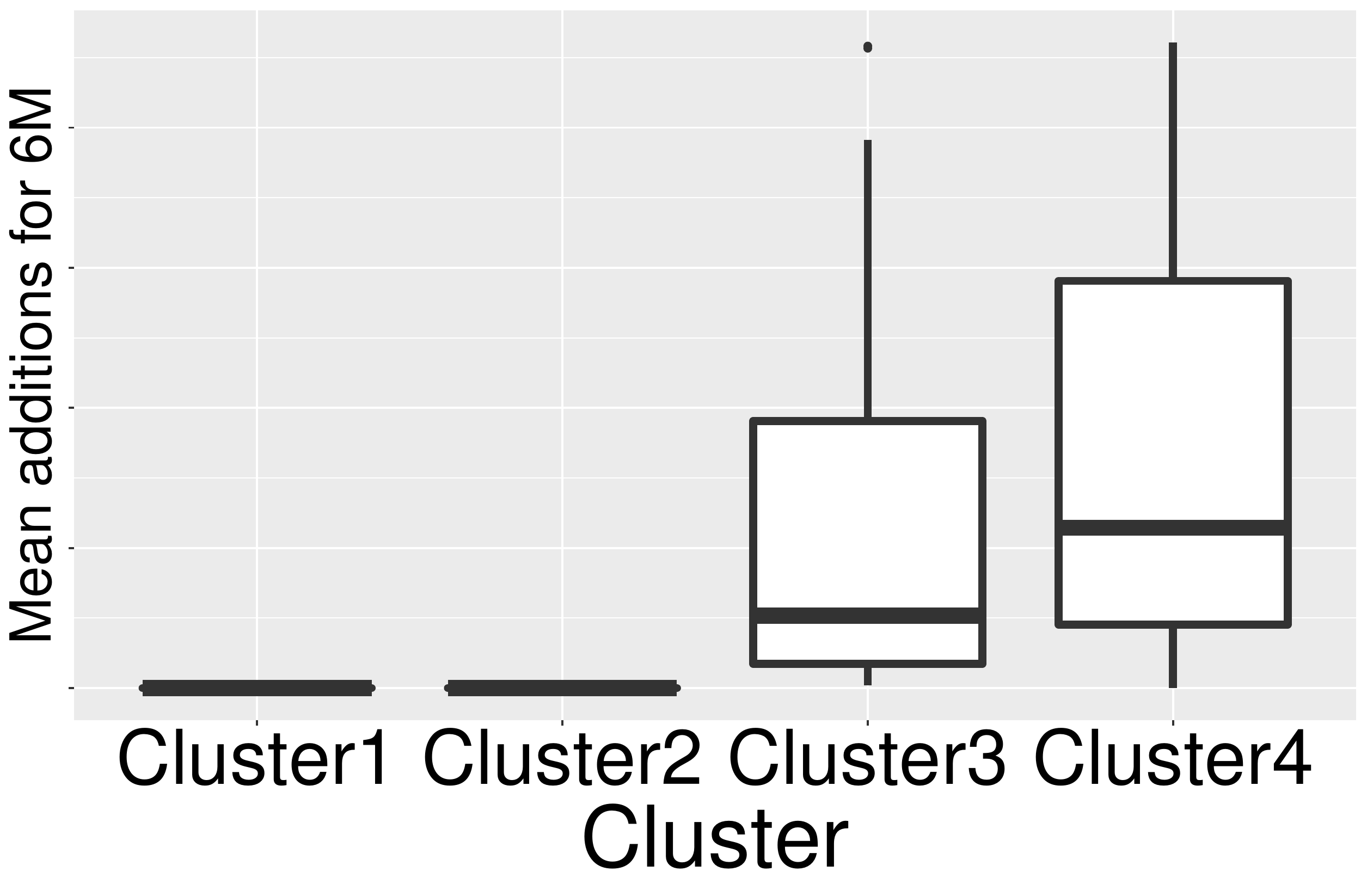}
    \end{minipage}\vfill
    \begin{minipage}[b]{0.31\textwidth}
        \centering \includegraphics[width=\linewidth]{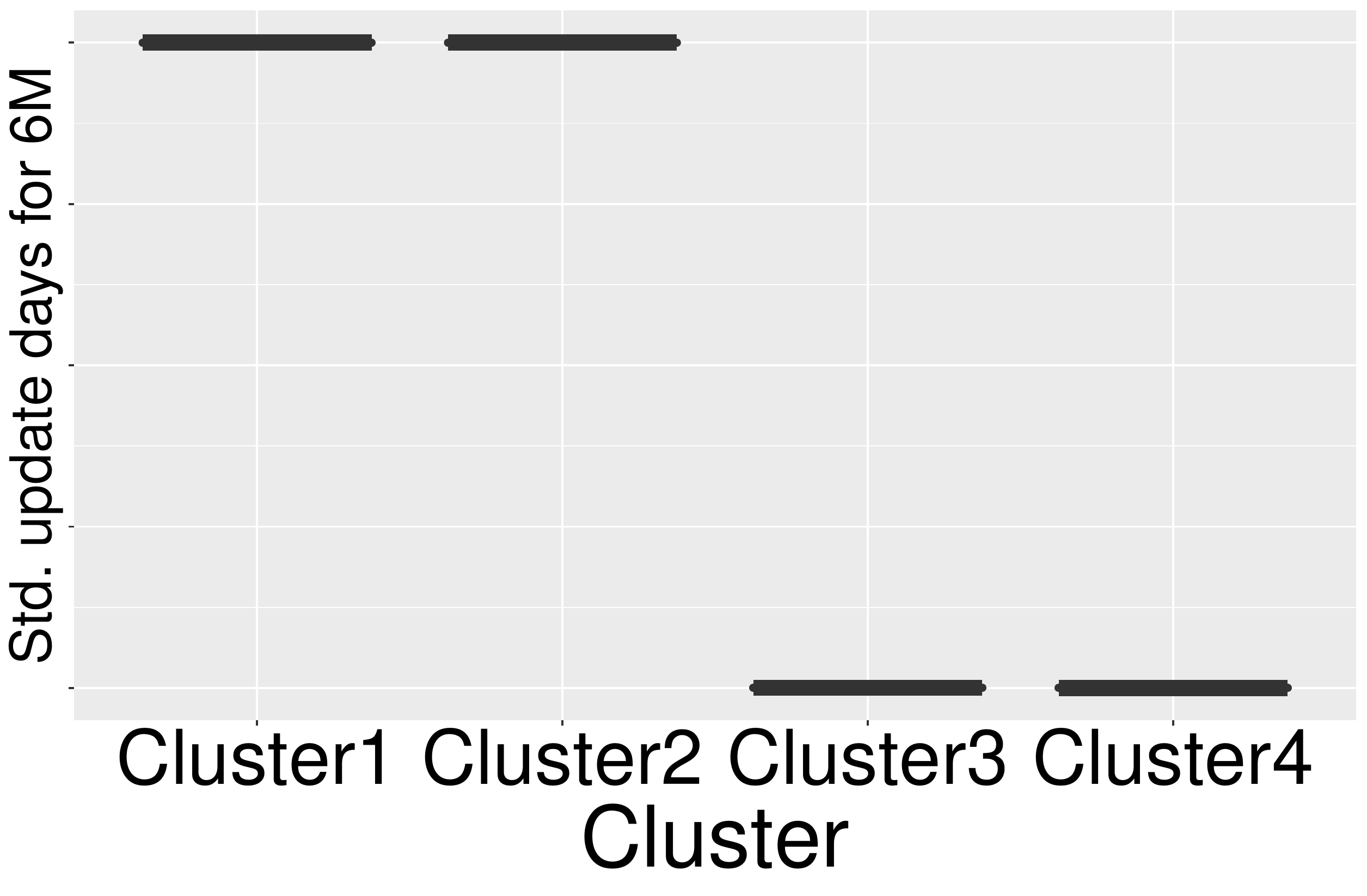}
    \end{minipage}\hfill
    \begin{minipage}[b]{0.31\textwidth}
        \centering \includegraphics[width=\linewidth]{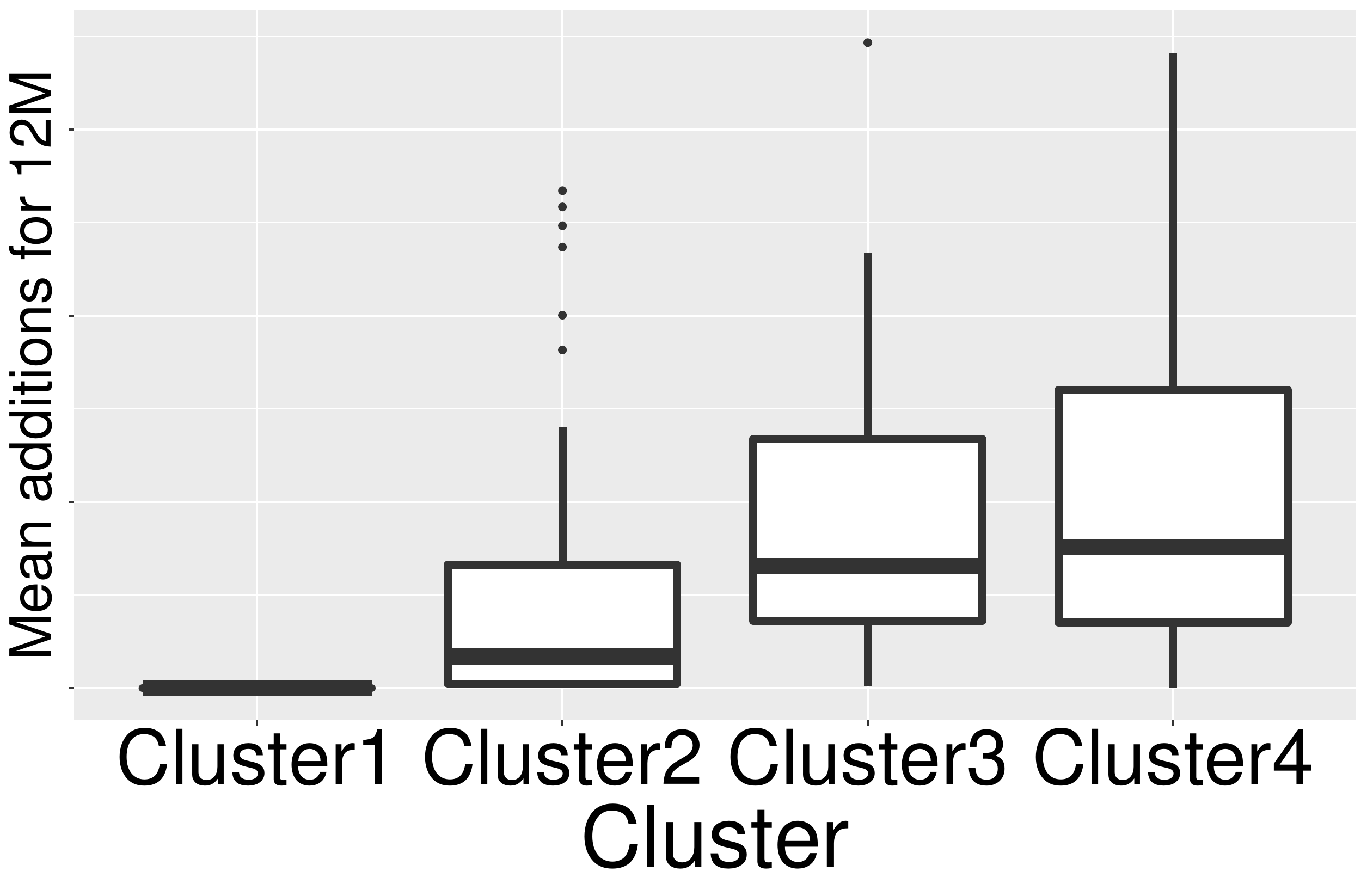}
    \end{minipage}\hfill
    \begin{minipage}[b]{0.31\textwidth}
        \centering \includegraphics[width=\linewidth]{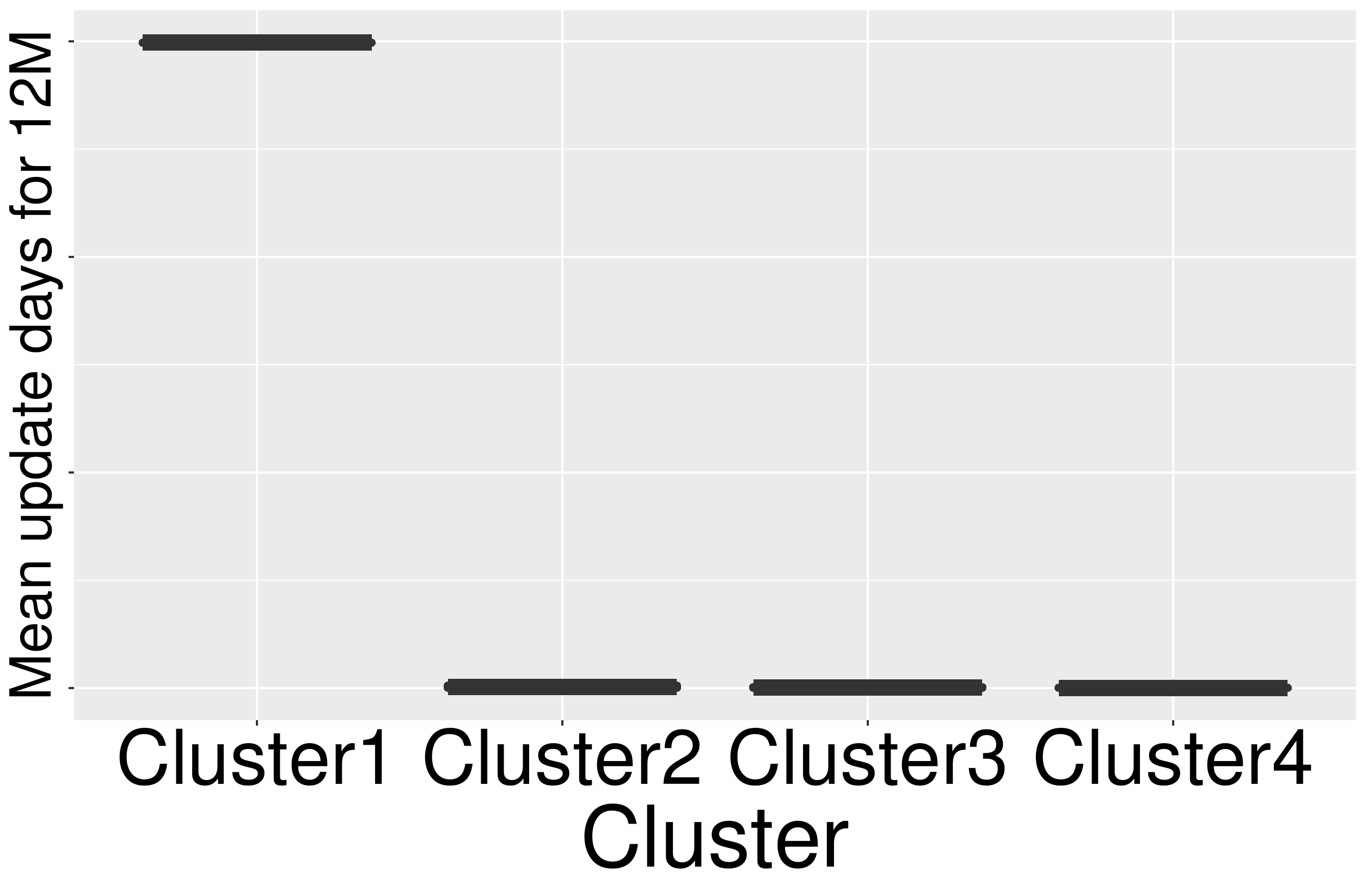}
    \end{minipage}\hspace{2em}\\ 
    \vspace{-0.2cm}
    \caption{Properties of each cluster in the top 6 key attributes (mean time interval between commits for 3M (M stands for months.), standard deviation time interval between commits for 3M, mean additions for 6M, standard deviation time interval between commits for 6M, mean additions for 12M, and mean time interval between commits for 12M). To better visualize the results, we exclude some outliers in analyzing mean additions for 6M and 12M.}
    \label{fig:features}
    \vspace{-0.2cm}
\end{figure}

We identified the clustering results' key attributes using the best-first search algorithm~\cite{pearl1984intelligent}. Fig.~\ref{fig:features} shows the comparison results of clusters for those key attributes (e.g., number of added lines of code, time interval between commits), indicating that the projects in Cluster 4 were persistently maintained well over time while the projects in Cluster 1, 2, and 3 were rarely updated for the last 12, 6, and 3 months, respectively. Surprisingly, 288 out of 592 (48.6\%) crypto projects (included in Cluster 1 and 2) have not been updated at all during the last six months, revealing that about half of crypto projects have been poorly maintained or abandoned.

To show the relevance of the clustering results, we compared the survivability of each cluster with the proportion of projects that disappeared from the market or were reported as a scam or inactive. We regard a project as inactive if it is not listed on CoinMarketCap, or its GitHub URL is unavailable (in September 2021). Note that a crypto project can only be listed on CoinMarketCap if it is available on a public exchange service and has a non-zero trading volume. Also, we regard a project as a scam if it is listed on Deadcoins~\cite{deadcoins} or Coinopsy~\cite{coinopsy}. The survivability results are presented in Table~\ref{tab:disappear_coins}, indicating that most crypto projects (59.1\%) disappeared from the market, or were reported as scams or inactive in just two years. Furthermore, it seems that the code update activities could be used as a signal for project survivability -- the cluster of frequently-updated crypto projects (in Cluster 4) has the lowest proportion (38.3\%) of inactive projects while the crypto projects in Cluster 1 (76.8\%), Cluster 2 (74.1\%), and Cluster 3 (62.3\%) that were rarely updated are highly like to be inactive or scams.

\begin{table}[t]
\vspace{-0.3cm}
    \centering
    \caption{Proportion of inactive/scam crypto projects in each cluster.}
    \resizebox{0.85\linewidth}{!}{
    \begin{tabular}{|c|r|r|r|r|r|r|}
        \hline
        ID & \multicolumn{1}{c|}{Number (\%)} & \multicolumn{1}{c|}{N/A in CoinMarketCap} & \multicolumn{1}{c|}{N/A in GitHub} & \multicolumn{1}{c|}{Coinopsy} & \multicolumn{1}{c|}{Deadcoins} & \multicolumn{1}{c|}{All} \\\hline
        1 & 207 (35.0\%) & 131 (63.3\%) & 7 (3.4\%) & 75 (36.2\%) & 30 (14.5\%) & 159 (76.8\%) \\\hline
        2 & 81 (13.7\%) & 52 (64.2\%) & 5 (6.2\%) & 18 (22.2\%) & 14 (17.3\%) & 60 (74.1\%) \\\hline
        3 & 61 (10.3\%) & 27 (44.3\%) & 3 (4.9\%) & 14 (23.0\%) & 15 (24.6\%) & 38 (62.3\%) \\\hline
        4 & 243 (41.0\%)& 70 (28.8\%) & 2 (0.8\%) & 35 (14.4\%) & 26 (10.7\%) & 93 (38.3\%) \\\hline
        All & 592 (100.0\%) & 280 (47.3\%) & 17 (2.9\%) & 142 (24.0\%) & 85 (14.4\%) & 350 (59.1\%) \\\hline
    \end{tabular}}
    \label{tab:disappear_coins}
    \vspace{-0.4cm}
\end{table}

\section{Security vulnerability analysis}
\label{sec:results_vulnerability}

This section presents the results of our security vulnerability analysis, which used the method described in Section~\ref{subsec:vulernability_method}. To gauge whether crypto project maintainers mitigate security vulnerabilities in a timely manner, we examined whether Bitcoin's known vulnerabilities were patched in 510 other C-based crypto projects. We first considered all known 39 Bitcoin CVEs since 2011\footnote{The Bitcoin CVEs before 2011 are not related to other crypto projects because the first altcoin was launched in the middle of 2011.}. However, we were only able to obtain code information about 11 CVEs from the Bitcoin Git messages (see the detailed information about those CVEs in Appendix~\ref{appendix:CVE}). For each of these CVEs, we got source code versions for before and after CVE occurrence, which enabled us to extract vulnerable codes and patched/fixed codes based on diff results. We manually verified the correctness of those codes. Appendix~\ref{appendix:Vulnerable and patched code fragment for CVE-2013-4627} shows an example of vulnerable and patched code fragments for ``CVE-2013-4627.'' 


\begin{table}[!ht]
\vspace{-0.6cm}
    \centering
    \caption{Proportion of vulnerable crypto projects.}
    \resizebox{0.56\linewidth}{!}{
    \begin{tabular}{|c|c|c|r|}
        \hline
        CVE ID & Type & CVSS & \multicolumn{1}{c|}{Number (\%)} \\\hline
        CVE-2012-1909 & DoS & 5.0 & 0 (0.0\%) \\\hline
        CVE-2012-3789 & DoS & 5.0 & 0 (0.0\%) \\\hline
        CVE-2012-1910 & DoS, Execute code & 7.5 & 0 (0.0\%) \\\hline
        CVE-2012-2459 & DoS & 5.0 & 0 (0.0\%) \\\hline
        CVE-2014-0160 & Overflow, Gain information & 5.0 & 149 (29.2\%) \\\hline
        CVE-2013-4627 & DoS & 5.0 & 294 (57.7\%) \\\hline
        CVE-2013-4165 & Gain information & 4.3 & 43 (8.4\%) \\\hline
        CVE-2014-0224 & Gain information & 5.8 & 176 (34.5\%) \\\hline
        CVE-2018-12356 & Execute code & 7.5 & 0 (0.0\%) \\\hline
        CVE-2018-17144 & DoS & 5.0 & 27 (5.3\%) \\\hline
        CVE-2019-6250 & Execute code, Overflow & 9.0 & 113 (22.2\%) \\\hline
    \end{tabular}}
    \label{tab:patch results}
\vspace{-0.5cm}
\end{table}

We determined whether 11 vulnerabilities were appropriately patched in each project based on such code fragment pairs. These results are summarized in Table~\ref{tab:patch results}, indicating that a significant number crypto projects leave some vulnerabilities unpatched. 294 (57.7\%) of cryptocurrencies are vulnerable to CVE-2013-4627~\cite{cve-2013-4627}, which has a Common Vulnerability Scoring System (CVSS)\footnote{CVSS provides a numerical (0-10) representation of the severity of an information security vulnerability.} score of 5.0. Thus, in many crypto projects, the memory of blockchain nodes can be exhausted with intentionally created dummy transactions. 176 (34.5\%) cryptocurrencies are vulnerable to CVE-2014-0224~\cite{cve-2014-0224}, named ``CCS Injection,'' that has a CVSS score of 5.8. This vulnerability originates from OpenSSL (\url{https://www.openssl.org/}), the most widely used open-source cryptographic library. This vulnerability allows attackers to carry out man-in-the-middle attacks for hijacking sessions and stealing sensitive information via a specially crafted TLS handshake. The impact of this vulnerability may be rather limited in crypto projects because SSL/TLS protocols are not directly used as communication protocols in most implementations. However, this shows that the code maintainaance -- and perhaps the security awareness level -- of a significant number of cryptocurrencies is poor because the patched OpenSSL version was released in June 2014. In the case of Bitcoin, this bug was immediately fixed by upgrading the OpenSSL library. 

As for real security risks, 149 crypto projects (29.2\%) are vulnerable to ``Heartbleed,'' CVE-2014-0160~\cite{cve-2014-0160}, which has a CVSS score of 5.0. As with CVE-2014-0224~\cite{cve-2014-0224}, this vulnerability originates from OpenSSL. 113 (22.2\%) are vulnerable to CVE-2019-6250~\cite{cve-2019-6250} that has a CVSS score of 9.0. This vulnerability originates from ZeroMQ libzmq (also known as 0MQ) (\url{https://github.com/zeromq/libzmq}), which is a library for high-performance asynchronous messaging. This vulnerability might be critical for some projects because a successful exploit could allow an attacker to overwrite an arbitrary amount of bytes beyond the bounds of a buffer, which, in turn, could be leveraged to run arbitrary code on a target system. The presence of these vulnerabilities demonstrates that a significant number of crypto projects are at actual risk of cyber-attacks; and people are investing in them without being aware of material security risks.



We note that the most serious vulnerabilities, namely CVE-2014-0224~\cite{cve-2014-0224}, CVE-2014-0160~\cite{cve-2014-0160}, and CVE-2019-6250~\cite{cve-2019-6250}, are vulnerabilities in third-party libraries imported into crypto projects. Given that the use of third-party libraries has become a common practice in the cryptocurrency industry, it is essential to keep them up to date. It is well known that they often contain serious bugs~\cite{Watanabe17:software}.

Table~\ref{tab:patch results} shows that no crypto projects containing the vulnerabilities (CVE-2012-1909~\cite{cve-2012-1909}, CVE-2012-3789~\cite{cve-2012-3789}, CVE-2012-1910~\cite{cve-2012-1910}, and CVE-2012-2459~\cite{cve-2012-2459}) which were discovered in 2012. We surmise that the reason for these results is that all but 24 of the projects we studied were forked after June 2012.

\begin{table}[t]
\centering
\caption{Proportion of inactive/scam crypto projects with the number of vulnerabilities.}
\label{table:delist_cve}
\resizebox{0.85\linewidth}{!}{
\begin{tabular}{|c|r|r|r|r|r|r|}
\hline
\# vuln. & \multicolumn{1}{c|}{Number (\%)} & \multicolumn{1}{c|}{N/A in CoinMarketCap} & \multicolumn{1}{c|}{N/A in GitHub} & \multicolumn{1}{c|}{Coinopsy} & \multicolumn{1}{c|}{Deadcoins} & \multicolumn{1}{c|}{All} \\\hline
0        &  99 (19.4\%) &  40 (40.4\%) & 5 (5.1\%) & 28 (28.3\%) & 17 (17.2\%) & 54 (54.5\%) \\\hline
$\geq 1$ & 411 (80.6\%) & 211 (51.3\%) & 8 (1.9\%) & 104 (25.3\%) & 64 (15.6\%) & 262 (63.7\%) \\\hline
$\geq 2$ & 227 (44.5\%) & 124 (54.6\%) & 5 (2.2\%) & 68 (30.0\%) & 36 (15.9\%) & 154 (67.8\%) \\\hline
$\geq 3$ & 138 (27.1\%) &  81 (58.7\%) & 1 (0.7\%) & 42 (30.4\%) & 23 (16.7\%) & 98 (71.0\%) \\\hline
$\geq 4$ &  25 (4.9\%)  &  19 (76.0\%) & 0 (0.0\%) & 7 (28.0\%) & 7 (28.0\%) & 21 (84.0\%)\\\hline
All      & 510 (100.0\%) & 251 (49.2\%) & 13 (2.5\%) & 132 (25.9\%) & 81 (15.9\%) & 316 (62.0\%) \\\hline
\end{tabular}}
\vspace{-0.5cm}
\end{table}

We also counted the number of vulnerabilities remaining in each crypto project. Table~\ref{table:delist_cve} lists the number of security vulnerabilities remaining for each of the 510 C-based projects, showing that 411 out of 510 (80.6\%) have at least one unpatched vulnerability.

To analyze whether the survivability of those altcoins was correlated with the number of vulnerabilities, we analyzed the proportion of inactive/scam projects with the number of vulnerabilities in the manner described in Section~\ref{sec:results_maintenance}. The results are presented in Table~\ref{table:delist_cve}, indicating that the number of vulnerabilities does indeed seem related to survivability. 84.0\% of altcoins with 4 or more vulnerabilities disappeared from the market or were reported as scams or inactive, while this fate befell only 54.5\% of those with no vulnerability.

\begin{figure}[!ht]
\vspace{-0.3cm}
\centering
\includegraphics[trim={0 0.1cm 0 0.8cm},clip, width=0.58\linewidth]{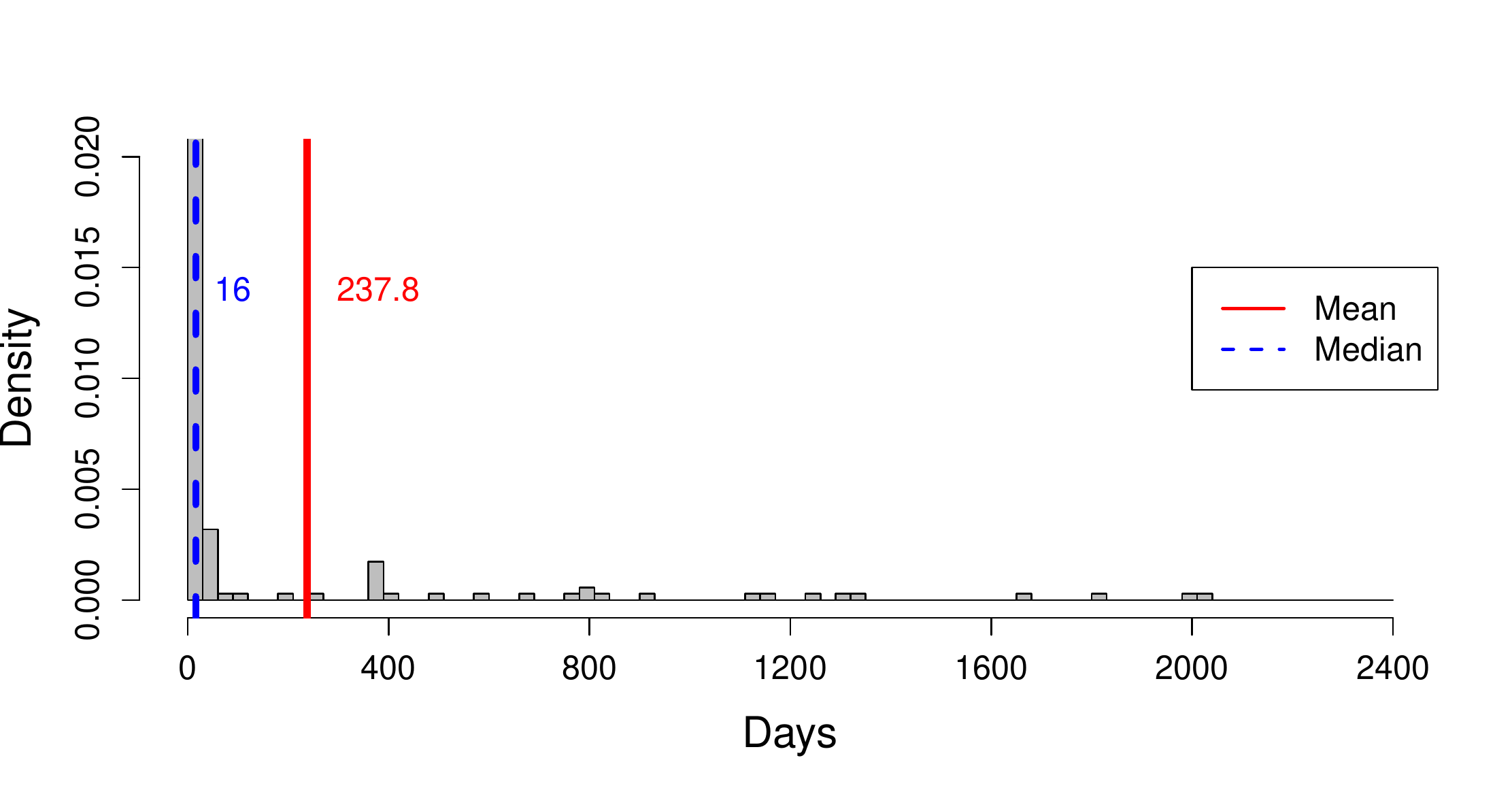}
\vspace{-0.4cm}
\caption{Number of days taken to fix a vulnerability.}
\label{fig:patch_days}
\vspace{-0.5cm}
\end{figure}

To analyze the time it took for the altcoin developers to fix each of the 11 vulnerabilities, we computed the elapsed time between (a) when Bitcoin first released a patch, and (b) when altcoins applied that patch on their code. Fig.~\ref{fig:patch_days} shows the median and mean values of the elapsed time, which were 16 and 237.8 days (with a standard deviation of 453.6 days). The large gap between median and mean values indicates that about half the vulnerabilities were patched quickly -- 56.6\% were patched within 16 days of the corresponding patch being applied to Bitcoin -- whereas the other half took weeks, months, or even years before they were fixed. 

To explore the relationship between code maintenance activities and unpatched  vulnerabilities, we compared the number of vulnerabilities remaining in each project between Clusters 1, 2, 3 and 4. In Cluster 1, the median and mean values are 2.0 and 1.80 (with a standard deviation of 1.20); in Cluster 2, the median and mean values are 1.5 and 1.73 (with a standard deviation of 1.30); in Cluster 3, the median and mean values are 1.0 and 1.28 (with ta standard deviation of 1.09); in Cluster 4, the median and mean values are 1.0 and 1.60 (with a standard deviation of 0.98). There is a statistically significant difference between the four clusters (p $<$ 0.0002, Kruskal-Wallis test).

\section{Code similarity analysis}
\label{sec:results_originality}

This section presents the results of the source code similarity analysis between crypto projects using the method described in Section~\ref{subsec:originality_method}.

\subsection{Analysis of similarity between altcoins and Bitcoin}
\label{subsec:newmethodsimilarity}

To investigate whether altcoins have made significant enhancements from their original parent project, we specifically focused on Bitcoin and its descendent cryptocurrencies because many altcoins were advertised as addressing various limitations of Bitcoin. This analysis is therefore crucial to evaluating not just the maintainability of crypto projects but also their justification and the truthfulness or otherwise of the claims used to market them. As living software should be continuously updated to add new features or fix bugs, projects that are not being maintained may have been quietly abandoned.

We used the heuristic methods described in Section~\ref{subsec:originality_method} to identify those crypto projects and estimate their forking times. When Heuristic 2 is used, we specifically set the similarity threshold to 92.9\% to determine whether a project is forked or not. This threshold was obtained by the mean similarity score (99.4\%) between Bitcoin and its altcoins minus three standard deviations (6.5\%), obtained from the Heuristic 1 method, based on the three-sigma limits. In Heuristic 2, if the similarity score between an altcoin and the version of Bitcoin used while creating the fork is greater than 92.9\%, we assumed that the altcoin was forked from Bitcoin even though the altcoin was not explicitly forked through GitHub. From 510 C-based crypto projects, we found that 157 altcoins are likely to have been forked from Bitcoin. Among those projects, we confirm that 154 projects have entered the mainnet; one project is still in the testnet phase; we cannot determine the status of the remaining two. 127 altcoins were discovered by Heuristic 1, while 30 altcoins were discovered by Heuristic 2. Table~\ref{table:altcoin_disappear} shows the proportion of those projects with their similarity scores, revealing that about a third of the 157 altcoins that are likely to have been forked from Bitcoin still have 90\% or more code similarity with the Bitcoin version used to spawn them. 

\begin{table}[t]
\vspace{-0.3cm}
\centering
\caption{Proportion of inactive/scam crypto projects with the similarity scores.}
\label{table:altcoin_disappear}
\resizebox{0.85\linewidth}{!}{
\begin{tabular}{|c|r|r|r|r|r|r|}
\hline
Similarity score & \multicolumn{1}{c|}{Number (\%)} & \multicolumn{1}{c|}{N/A in CoinMarketCap} & \multicolumn{1}{c|}{N/A in GitHub} & \multicolumn{1}{c|}{Coinopsy} & \multicolumn{1}{c|}{Deadcoins} & \multicolumn{1}{c|}{All} \\\hline
$\geq 95\%$ & 34 (21.7\%) & 16 (47.1\%) & 0 (0.0\%) & 12 (35.3\%) & 11 (32.4\%) & 24 (70.6\%) \\\hline
$\geq 90\%$ & 53 (33.8\%) & 25 (47.2\%) & 0 (0.0\%) & 17 (32.1\%) & 17 (32.1\%) & 38 (71.7\%) \\\hline
$\geq 80\%$ & 66 (42.0\%) & 31 (47.0\%) & 1 (1.5\%) & 19 (28.8\%) & 17 (25.8\%) & 46 (69.7\%) \\\hline
$\geq 60\%$ & 95 (60.5\%) & 36 (37.9\%) & 1 (1.1\%) & 24 (25.3\%) & 22 (23.2\%) & 57 (60.0\%) \\\hline
$<50\%$     & 43 (27.4\%) & 11 (25.6\%) & 1 (2.3\%) & 10 (23.3\%) & 5 (11.6\%) & 18 (41.9\%) \\\hline
All         & 157 (100.0\%) & 52 (33.8\%) & 4 (2.5\%) & 36 (22.9\%) & 29 (18.5\%) & 82 (52.2\%) \\\hline
\end{tabular}}
\vspace{-0.6cm}
\end{table}

To verify the correctness of the forked projects identified by our heuristic methods, we compare those projects with a list of known Bitcoin descendants (\url{https://mapofcoins.com/bitcoin/}). From the 175 cryptocurrencies active in December 2017, we selected 68 included in our dataset for testing (hereinafter referred to as the ``known list''). Most of the remaining ones were defunct when we started collecting data on July 22nd, 2018. For a fair comparison, among the 157 cryptocurrencies identified as forked projects by our method, we also considered 139 (hereinafter referred to as ``our list'') that were forked before December 2017. By comparing ``our list'' against the ``known list,'' we observed only 33 cryptocurrencies in both lists. That is, our method was able to identify additional 106 coins whose Bitcoin ancestry was not included in the ``known list.'' However, our method failed to infer the forking relationships of the 35 remaining coins in that list. We manually checked these failure cases and found that those altcoins were created by externally uploading an old version of Bitcoin as new files to a repository. Although Heuristic 2 is designed to detect such cases, Heuristic 2 can also suffer false negatives because, for performance reasons, it does not examine all previous versions of Bitcoin. This limitation is discussed in Section~\ref{sec:limit}.



\begin{figure}[t]
\vspace{-0.1cm}
\centering
\includegraphics[width=0.58\linewidth]{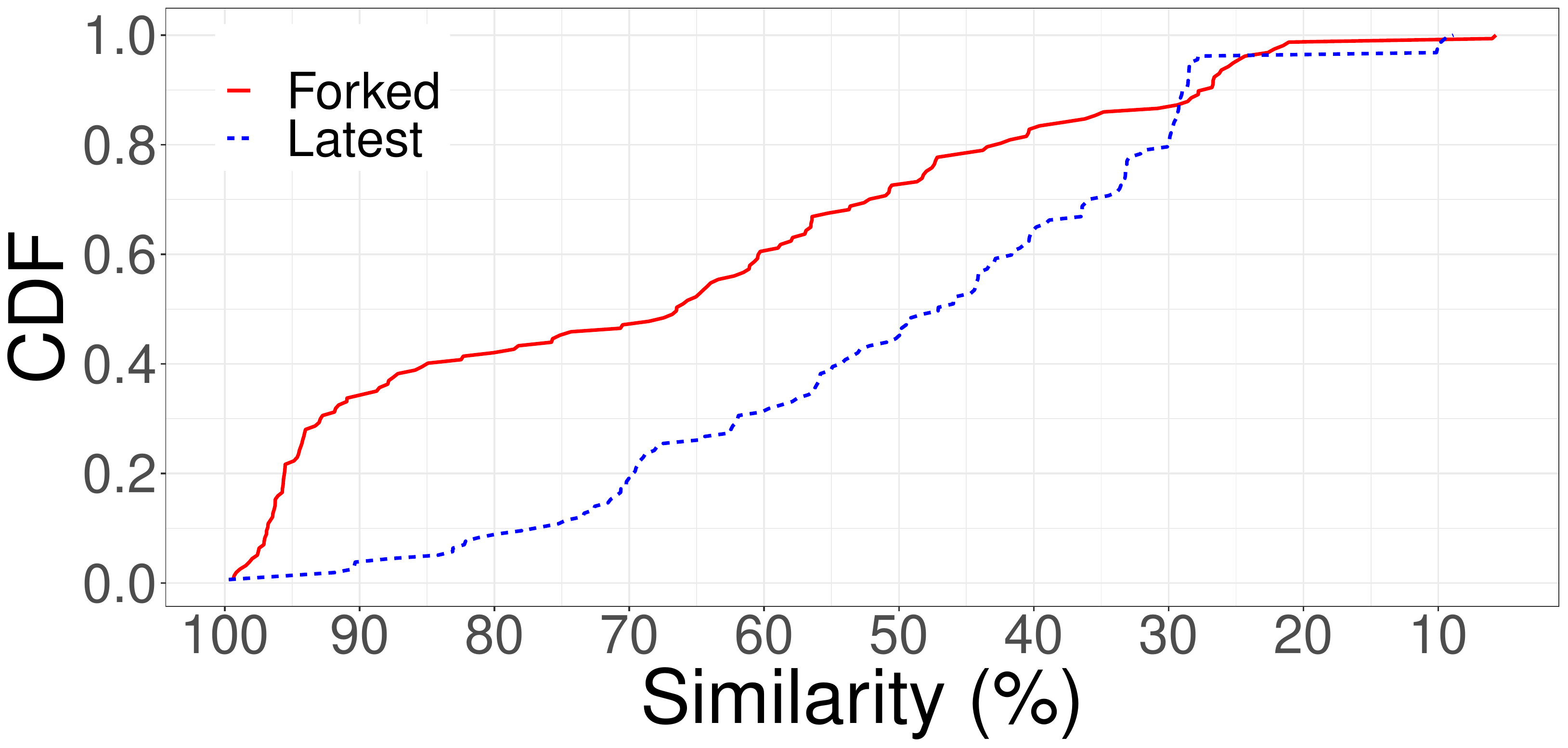}
\vspace{-0.2cm}
\caption{CDFs of the similarity scores between altcoins and Bitcoin (Latest: similarity scores between the latest version of altcoins and the latest version of Bitcoin, Forked: similarity scores between the latest version of altcoins and the version of Bitcoin used for the fork).}
\label{fig:difference_c}
\vspace{-0.6cm}
\end{figure}

Fig.~\ref{fig:difference_c} shows the CDF (dashed blue line) of the similarity scores between Bitcoin and its descendent cryptocurrencies. Only 6 out of 157 (3.8\%) crypto projects forked from Bitcoin have 90\% or more code similarity with the latest version of Bitcoin. From these findings, however, we cannot conclude how much altcoins have changed over time because Bitcoin itself changes significantly. Therefore, to examine how much a forked altcoin has changed from Bitcoin, we compute the similarity score between the latest source code of the forked project and the source code of Bitcoin at the time of the fork. Fig.~\ref{fig:difference_c} shows the CDF (solid red line) of the similarity scores between them. Among 157 projects forked from Bitcoin, 21.7\% of the pairs have 95\% or more code similarity; 33.8\% of the pairs have 90\% or more; 42.0\% of the pairs have 80\% or more; and 60.5\% of the pairs have 60\% or more. These results confirm that forked projects are most similar to their parent version of Bitcoin. This finding implies that a significant number of forked projects contain only negligible changes from the parent version of Bitcoin at the time of the fork.


To show the relevance of code similarity, we analyzed the relationship between the survivability of the projects forked from Bitcoin and their similarity scores. As presented in Table~\ref{table:altcoin_disappear}, the projects with high similarity scores are likely to disappear from the market or be reported as a scam or inactive: 70.6\% of altcoins with 95\% or more code similarity suffered this fate, compared with only 41.9\% of altcoins with similarity scores less than 50\%.



\subsection{Case studies}
\label{subsec:Case studies}

To identify altcoins that did not implement the unique features claimed in their whitepapers, we examined source code and whitepaper contents for the top 20 altcoins with the highest similarity scores with their parent version of Bitcoin (see Appendix~\ref{appendix:Top 20 altcoins that are highly similar to the version of Bitcoin used for the fork}).




Among those altcoins, MinCoin (MNC), GapCoin (GAP), Litecoin (LTC)\footnote{Although Litecoin was initially forked from Bitcoin on October 13th, 2011, the old project was deprecated, and a new Litecoin project was created again by forking from the ``August 13th, 2018'' version of Bitcoin.}, Platincoin (PLC), Florincoin (FLO), MinexCoin (MNX), Riecoin (RIC), Fujicoin (FJC), Feathercoin (FTC), Octocoin (888), and Uniform Fiscal Object (UFO) are variants of Bitcoin that made slight modifications in the way cryptographic algorithms are used for proof of work, and in policy parameter values (e.g., total supply, block generation time). Uniform Fiscal Object has been frequently updated -- every 0.1 days on average. This project has been clustered into Cluster 4, consisting of frequently updated projects. However, most commits we inspected contain changes in comments, icons, genesis block, and total supply. We found the only meaningful change was the replacement of SHA-256 with NeoScrypt in the hash function used for mining.

\begin{figure}[!ht]
\vspace{-0.2cm}
\centering
\includegraphics[width=0.99\linewidth]{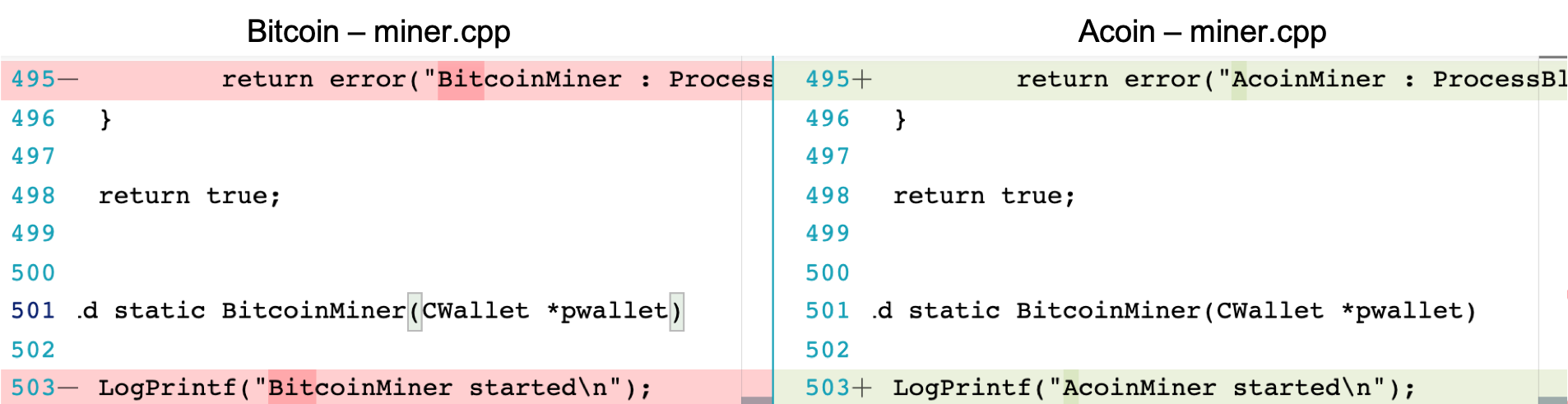}
\vspace{-0.3cm}
\caption{A change in the ``miner.cpp'' file in Acoin.}
\label{fig:code_example}
\vspace{-0.4cm}
\end{figure}

We found that the remaining nine crypto projects do not contain any technically meaningful code change. The most frequent changes were cryptocurrency name changes. For example, Acoin (\url{https://aco.ample-cosplay.com/}) is a copycat of Bitcoin that did not add any noticeable new feature after the fork. The only changes we noticed were replacing ``bitcoin'' with ``acoin'' in some code fragments (see Fig.~\ref{fig:code_example}). 


More worryingly, six crypto projects did not appropriately implement features promised in their whitepapers. For example, according to the whitepaper (\url{http://bit.ly/blockarray-whitepaper}) written for Block Array (ARY), it was designed to facilitate the development of protocols for logistics. The whitepaper also presented a timeline and road map until 2020. However, we found just two commits for Block Array: the first was about uploading the Bitcoin project files, and the second, on December 29th, 2017, created the \texttt{README.md} file. Although Block Array reused the entire Bitcoin source code, the whitepaper does not explain this. Its official website (\url{https://www.blockarray.com/}) does not even exist. 






There were seven projects for which the official websites have disappeared. For example, PLNcoin's (PLNC) official website URL (\url{https://plncoin.org/}) has been redirected to a suspicious website. CHIPS' Git repository only contains the link of the official Bitcoin Core website. 

Finally, sixteen projects disappeared from the market, or were reported as scams or inactive by September 2021. Half of them have already been reported as scam coins: GapCoin, Freicoin, PLNcoin, MinexCoin, Riecoin, SecureCoin, and OctoCoin were flagged as scams in Coinopsy~\cite{coinopsy}, while Betacoin, Freicoin, PLNcoin, MinexCoin, Riecoin, FujiCoin, SecureCoin, and Uniform Fiscal Object were flagged as scams in Deadcoins~\cite{deadcoins}. Our findings indicate that code-similarity metrics might provide a better signal to detect inactive or scam projects than such public reporting services. 

\section{Discussion}
\label{sec:discuss}

\subsection{Is the cryptocurrency market a market for lemons?}
\label{subsec:How can we fix this}

There are strong financial incentives for individuals and companies to copy source code from existing crypto projects as they can launch a project quickly and cash in from an ICO. Many projects advertise fancy whitepapers without actually delivering promised features. Cohney et al.~\cite{cohney2019coin} found that smart contracts and ICO disclosures often do not match. Some popular ICOs have retained the power to modify their tokens' rights but did not disclose that ability in their whitepaper, which is consistent in our findings. We discovered some project teams (e.g., Block Array and Acoin) that focused on creating fancy whitepapers and websites rather than implementing promised features. Naive investors may be unable to distinguish such fly-by-night projects from well-maintained ones. 

To analyze the relationship between price and code quality, we computed Pearson's correlation coefficients between the market capitalization (i.e., price $\times$ the number of coins available) of crypto projects and their code maintenance activities, security, and code originality, respectively. The results show that there was no correlation between market capitalization and code maintenance activities ($r = 0.05$, p $=$ 0.44). Similarly, we failed to show correlation between market capitalization and the code similarity to Bitcoin ($r = 0.11$, p $=$ 0.29). However, the number of security vulnerabilities had a weak negative correlation with market capitalization ($r = -0.14$, p $<$ 0.05). These results indicate that altcoin code quality may not affect market prices: i.e., poor code quality does not necessarily mean that an altcoin is cheap, and vice versa. 

This suggests that the crypto market has turned into a classic ``market for lemons''~\cite{Akerlof70:lemon}: in the presence of asymmetric information and unconstrained opportunism, adverse selection can lead to inefficient markets where producers keep making low-quality products (lemons) instead of high-quality ones (peaches). In the crypto world, non-tech-savvy consumers will probably just read whitepapers and invest in a project without knowing how actively it is being maintained. The inevitable outcome is a crypto market full of scammy projects.

\subsection{Limitations}
\label{sec:limit}

\textbf{Threats to validity.} 
One threat to external validity is whether we used representative crypto projects for our study. A bias could have been introduced since we only analyzed C-language projects for vulnerability and originality. However, we note that most of the collected crypto projects, 86.3\%, have been developed in C/C++. A primary threat to internal validity is the bias that may have been introduced by researchers while conducting the case studies (see Section~\ref{subsec:Case studies}). We had to go through the whitepapers manually to identify altcoins that did not implement promised features (listed in the whitepapers). To minimize bias, the three researchers read the whitepapers separately, identified feature claims, and inspected the source code for compliance. After independently performing these analyses, the three researchers discussed their observations until they reached a consensus. A primary threat to construct validity is the metrics used in the study. To evaluate the crypto project maintenance quality, we used the 32 features available on GitHub. To evaluate their security, we examined patches implemented for known security vulnerabilities. To evaluate their originality, we measured the code similarity between the forked and original projects. We note that these metrics are just one of the few that can be used for evaluation, and more metrics might be studied in the future to improve the precision of our analysis.

\noindent \textbf{Determining forking points.} We developed two heuristics to guess the fork date of a given project (see Section~\ref{subsec:originality_method}). However, our methods can fail to estimate the exact forking time of a forked project, or to infer the forking relationship, when an old version of Bitcoin is used as new files to a repository instead of using the fork feature provided by GitHub. To tackle this issue, one might make comparisons with all past versions of the original project. 

\noindent \textbf{Analyzing maintenance activities with Git messages.} When we analyze the maintenance activities on crypto projects, we only used the information obtained from Git. However, this information can be fabricated intentionally with dummy commits, dummy contributors, etc. We found that 22 projects in Cluster 4 are highly similar to their original projects (i.e., their similarity score is greater than or equal to 90\%) although they have been frequently updated. In such projects, we believe that developers may upload meaningless code periodically and add unnecessary comments so that their project appears to be on schedule and going well. We therefore suggest that further measurement techniques (including analysis of security vulnerabilities and code originality) should be developed as indicators of crypto project maintenance.

\section{Related work}
\label{sec:related}



\noindent
\textbf{Understanding the landscape of cryptocurrencies.}
As cryptocurrencies became popular, we have seen a broader academic interest in the activities associated with blockchains such as user privacy~\cite{Biryukov14:bitcoin,Goldfeder18:bitcoin}, market manipulation~\cite{xu2019anatomy}, money laundering~\cite{brenig2015economic,barone2019cryptocurrency}, network availability~\cite{vasek2014empirical}, bug characteristics~\cite{wan2017bug}, and scams~\cite{vasek2015there,vasek2018analyzing}. Zetzsche et al.~\cite{zetzsche2017ico} showed that many ICOs have utterly inadequate information disclosure; e.g., more than half of ICO whitepapers are silent on the initiators. Oliva et al.~\cite{oliva2020exploratory} conducted a study of smart contracts in the Ethereum platform and found that only a small number (0.05\%) of the smart contracts were actively used in most (80\%) of the transactions that were sent to contracts. Trockman et al.~\cite{trockman2019striking} investigated if there is a dynamic relation between software metrics and market capitalization for 268 cryptocurrencies using dynamic regression models and found that there was no compelling evidence of a relationship. Lucchini et al.~\cite{Lucchini04:cryptocurrency} found that the number of developers working on a cryptocurrency project correlates positively with its market capitalization. This paper further analyzed correlation coefficients between market capitalization and code maintenance activities, unpatched security vulnerabilities, and code similarity, respectively. Reibel et al.~\cite{reibel2019short} analyzed the code diversity of cryptocurrencies' open-source code repositories. Their study is similar to ours in some ways because both exploit project information extracted from repositories to evaluate the originality of crypto projects. However, their approach relied on rather naive rules (name derivations, copyright derivations, common commits, and file derivations) that often misinterpret the relationships between crypto projects. In addition to offering a more thorough code-diversity analysis, we analyze security vulnerabilities and measure how much Bitcoin's descendent projects differ from Bitcoin. 




\noindent
\textbf{Evaluating the quality of project management.} Many techniques have been proposed to evaluate the quality of open-source projects. Thung et al.~\cite{thung2013network} constructed social graphs between projects to identify which were influential by
using PageRank on the graph. Hu et al.~\cite{hu2016influence} also measured the importance
of software repositories using social-network analysis techniques. In those studies, social graphs between projects were constructed by using project information explicitly provided by Git, such as common developers between projects or star relationships.  Unlike their work, our proposal can identify hidden fork relationships between projects because we assume that some developers may intentionally cover up a fork relationship. Coelho et al.~\cite{Coelho20:Software} proposed a method to evaluate the maintenance of open-source projects using classification algorithms and 13 features such as the number of commits in the last time interval and the maximum number of days without commits in the last time interval. We extended their analysis method with 19 additional features such as commit time interval and MDE value to evaluate crypto project maintenance activities.




\noindent
\textbf{Detecting known vulnerabilities.} Several proposals (e.g., \cite{jiang2007deckard,sajnani2016sourcerercc,perl2015vccfinder}) have been introduced for detecting similar vulnerabilities in code bases. However, their detailed inspection techniques are too time-consuming for mass measurement. The total number of commits analyzed in our experiments is over 1.6 million. Our study's precondition -- identifying bugs introduced by the parent project (e.g., Bitcoin) in a child project (e.g., Litecoin) -- makes our evaluation tractable even without detailed inspection techniques. Li et al.~\cite{li2017large} collected security patches using Git commits and conducted a large-scale empirical study of them. However, unlike Li et al.'s work~\cite{li2017large}, we identified the locations of those vulnerabilities and patches in Bitcoin derivatives using their source code as well as Git commits because we wanted to also cover the cases where patch information is not reported through Git commits. Hum et al.~\cite{hum2020coinwatch} used clone-code detection to find vulnerable crypto projects, and discovered 786 vulnerabilities from 384 projects using 4 CVEs (CVE-2018-17144, CVE-2016-10724, CVE-2016-10725, and CVE-2019-7167). In this paper, we report a much more efficient vulnerable code detection method: we use code commits to inspect code changes only, rather than examining the full source code. This enabled us to cover all previous versions of crypto projects and identify when security vulnerabilities were patched. We also used 11 CVEs -- CVE-2018-17144 is only common CVE between Hum et al.'s work and ours.

\section{Conclusion}
\label{sec:conclusion}

Many altcoins have been abandoned by their developers, or were outright scams from the start. Yet people continue to trade them, to use them, and even to invest in them. To investigate abuses that happen at scale, we also need analysis tools at scale. Therefore we built a semi-automated framework to inspect the maintenance and security of altcoins by analyzing their source code and project activities.


Our analyses indicate that many projects are not maintained properly or at all; that many leave known vulnerabilities unpatched; and that many do not even implement features promised in their whitepapers. About half of the 592 projects we inspected have not been updated during the last six months.  About a third of the projects forked from Bitcoin have 90\% or more code similarity with the Bitcoin version used to spawn them. Most concerning, over 80\% of 510 crypto projects we analyzed have at least one unpatched vulnerability. 

If the altcoin market is to become less of a lemons market, market participants need better ways of comparing project originality, liveness, and security. We believe that systematic code analysis reports are a necessary first step.

\bibliographystyle{splncs04}
\bibliography{main}

\appendix
\clearpage
\section{Features for evaluating maintenance efforts}
\label{sec: Features used for evaluating maintenance efforts}
Table~\ref{table:features} summarizes the list of the features that can be collected from a GitHub project and used for evaluating maintenance efforts. We categorize the 32 features into 3 groups by their characteristics.

\begin{table*}[!h]
\vspace{-0.5cm}
\begin{center}
\caption{List of maintenance features. We set $k=3$, $6$, and $12$ months for the features for code updates to examine short-, mid-, long-term update activities, respectively, on projects.}
\label{table:features}
\resizebox{0.99\linewidth}{!}{
\begin{tabular}{|p{0.2\linewidth}|p{0.28\linewidth}|p{0.84\linewidth}|}
\hline
\multicolumn{1}{|c|}{Group} & \multicolumn{1}{|c|}{Feature} & Description \\ \hline
\multirow{15}{\linewidth}{5 features for developers' engagement level ($k=$3, 6 and 12 period)} & \multirow{3}{\linewidth}{Commits} & Commits are easily one of the most frequented activities by a developer using GitHub. A commit is like `saving' an updated file to its original folder. \\ \cline{2-3}
& \multirow{3}{\linewidth}{Branches} & A branch is a parallel version of a repository. It is contained within the repository, but does not affect the primary or master branch allowing you to work freely without disrupting the ``live'' version. \\ \cline{2-3} 
& \multirow{2}{\linewidth}{Releases} & Releases are GitHub's way of packaging and providing software to the users. \\ \cline{2-3} 
& \multirow{2}{\linewidth}{Contributors} & A contributor is someone who has contributed to a project by having a pull request merged but does not have collaborator access. \\ \cline{2-3} 
& \multirow{3}{\linewidth}{Pull requests} & Pull request are proposed changes to a repository submitted by a user and accepted or rejected by a repository's collaborators. Like issues, pull requests each have their own discussion forum.  \\ \cline{2-3}
& \multirow{2}{\linewidth}{MDE ($n=k$)} & MDE ($n=k$) is the mean developer engagement, which indicates the mean number of contributors each $k$ period within 12 months.             \\ \hline
\multirow{13}{\linewidth}{6 features for popularity} & \multirow{2}{\linewidth}{Watch} & ``Watch a project'' means the user are notified whenever there are any updates. Watch is the number of the users who set watching project. \\ \cline{2-3} 
& \multirow{3}{\linewidth}{Star} & Starring makes it easy to find a repository or topic again later. The user can see all the repositories and topics you have starred by going to your stars page. \\ \cline{2-3} 
& \multirow{3}{\linewidth}{Fork} & A `fork' is a personal copy of another user's repository that lives on your GitHub account. Forks allow the user to freely make changes to a project without affecting the original. \\ \cline{2-3} 
& \multirow{3}{\linewidth}{Issues} & Issues are suggested improvements, tasks or questions related to the repository. Issues can be created by anyone and are moderated by repository collaborators. \\ \cline{2-3} 
& \multirow{2}{\linewidth}{Open Issues} & Open issue is the number of the created issues. If the developer decides the issue is solved, the issue will be closed. \\ \cline{2-3} 
& Closed Issues & Closed issue is the number of the solved issues. \\ \hline
\multirow{14}{\linewidth}{21 features for code updates ($k=$3, 6, and 12 months)} & Mean additions for $k$ months & \multirow{2}{\linewidth}{Mean additions for $k$ months is the mean number of added lines which are recorded for last $k$ months.} \\ \cline{2-3}
& Std. additions for $k$ months & Std. additions for $k$ months is the standard deviation of added lines which are recorded for last $k$ months. \\ \cline{2-3} 
& Mean deletes for $k$ months & \multirow{2}{\linewidth}{Mean deletions for $k$ months is the mean number of deleted lines which are recorded for last $k$ months.} \\ \cline{2-3} 
& Std. deletes for $k$ months & Std. deletions for $k$ months is the standard deviation of deleted lines which are recorded for last $k$ months. \\ \cline{2-3} 
& Mean time interval between commits for $k$ months  & \multirow{3}{\linewidth}{Mean time interval between commits for $k$ months is the mean update time between commits which are recorded for last $k$ months.} \\ \cline{2-3} 
& Std. time interval between commits for $k$ months  & \multirow{3}{\linewidth}{Std. time interval between commits for $k$ months is the standard deviation of update time between commits which are recorded for last $k$ months.}  \\ \hline
\end{tabular}
}
\end{center}
\vspace{-0.5cm}
\end{table*}

Unlike the other features for developers' engagement level, MDE cannot be directly obtained from GitHub but can be calculated with the number of contributors during a certain time period. MDE was introduced as a measurement of how effective (on average over time) an open-source project is at making use of its human resources~\cite{spinellis2009evaluating}. It is calculated by Equation~\eqref{eq:mde} where $n$ is the number of periods and $contributor_{i}$ is the number of non-duplicated contributors who made commits in the $i$th time period. For example, if the number of contributors is 6 for the first time period, 3 for the second time period, 6 for the last time period, and 12 for the total periods after de-duplicating, MDE becomes 0.41 since $\frac{6 / 12 + 3 / 12 + 6 / 12}{3} \approx 0.41$. 

\begin{equation} \label{eq:mde}
MDE = \frac{\sum\limits_{i=1}^{n} (contributor_{i} / \sum\limits_{j=1}^{n} contributor_{j})}{n} 
\end{equation}

\section{Process of security vulnerability analysis}
\label{appendix:Process of security vulnerability analysis}

Our security vulnerability process consists of two steps as follows.

\begin{enumerate}
    \item From Git, we download the latest source code for a crypto project to be inspected, remove all whitespace, and use a string match algorithm to find a piece of code associated with a given Common Vulnerabilities and Exposure (CVE). If we find a match, we mark that project as vulnerable to that specific CVE and terminate the process -- this existence of vulnerable code indicates that the project has not been patched to mitigate that CVE. If there is no match, we conclude that the CVE never existed or has already been patched. Therefore, we move to Step~\ref{step:2} to find the times at which the vulnerable and patched codes were added, respectively.
    
    \item \label{step:2} We obtain the information about added and deleted lines of code from the project repository on GitHub. We first check whether the vulnerable code was included using the diff result of source code before and after a commit. However, GitHub does not provide code change information if the number of changed files is greater than 30. Therefore, in such cases, we download the full source code for the commit and check whether the source code is vulnerable to CVEs. If we find a vulnerable piece of code, we log the time at which this vulnerable code was added and search for the corresponding security patch in the same manner. If we find a security patch, we also log the time at which the patch was added. However, if we do not find a vulnerable piece of code in a repository, we assume that the repository has been forked from the Bitcoin project where vulnerabilities have been fixed, or the code related to the CVEs has not been included in the project from its beginning to the present.
\end{enumerate} 

We note that the proposed technique is more efficient than finding a piece of code in all previous versions of an altcoin's source code. For example, in the case of Carboncoin, when we downloaded and inspected all 34 previous versions of the Carboncoin source code, it took about 18.1 seconds. However, our proposed technique took only about 0.024 seconds which is about 754.2 times more efficient than the naive method.

\section{Selection of $k$ in $k$-means clustering}
\label{appendix:Selection of k in k-means clustering}

To implement the $k$-means clustering algorithm, we used Weka~\cite{weka}, which is a widely used machine learning platform. The silhouette coefficient can be used to interpret and validate consistency within clusters of data~\cite{rousseeuw1987silhouettes}. The Silhouette coefficient is computed by Equation~\eqref{eq:silhouette}, where $a(i)$ is the mean distance of cluster $i$ and all other data points in the same cluster, and $b(i)$ is the smallest mean distance of cluster $i$ to all points in any other cluster to which $i$ does not belong. If the coefficient value is close to 1, the clusters are well-clustered.

\begin{equation} \label{eq:silhouette}
Silhouette_{k} = \frac{\sum\limits_{i=1}^{k} \frac{b(i)-a(i)}{max(a(i),b(i))}}{k}
\end{equation}

\begin{figure}[!h]
\vspace{-0.2cm}
\begin{center}
\begin{tabular}{c}
\includegraphics[trim={0 0.4cm 0 0},clip, scale=0.26]{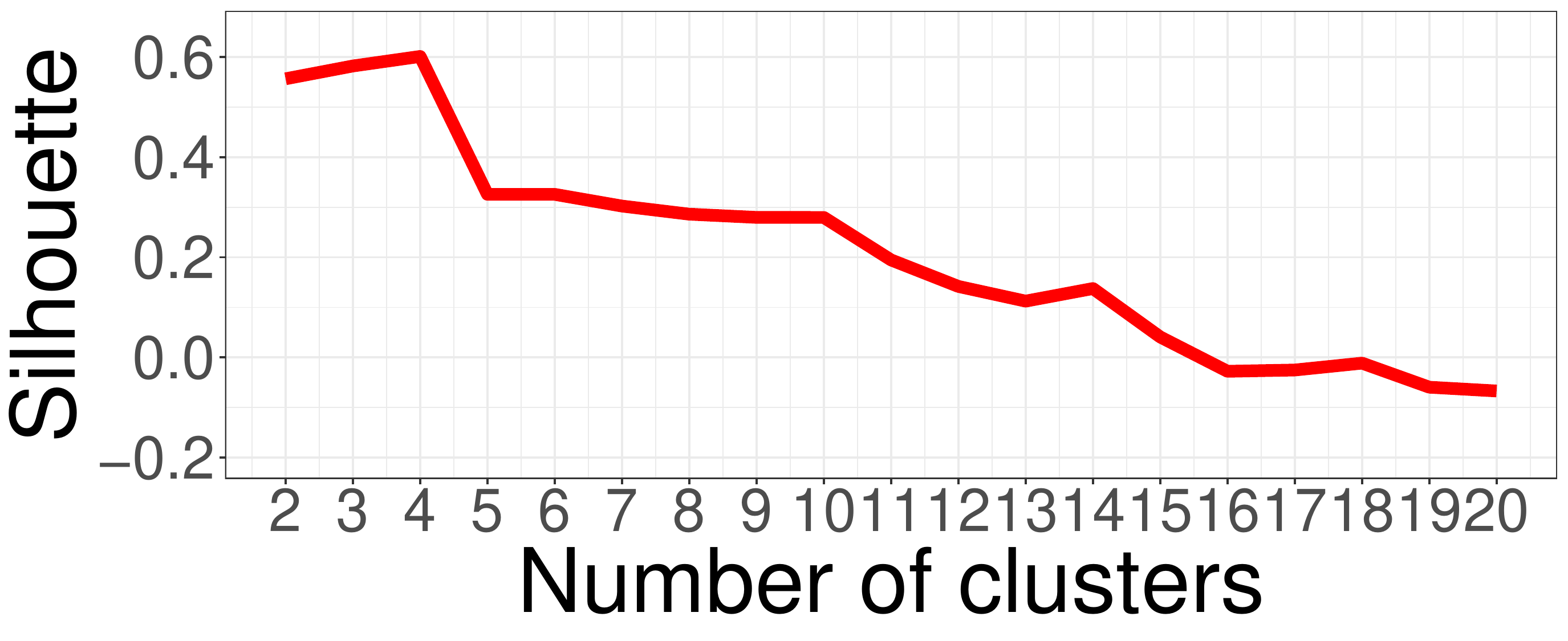} %
\end{tabular}
\end{center}
\vspace{-0.4cm}
\caption{Silhouette results from $k=2$ to $20$.}
\label{fig:silhouette}
\vspace{-0.3cm}
\end{figure}

To determine the optimal number of clusters, we computed the Silhouette coefficients for clustering results varying $k$ from 2 to 20. Fig.~\ref{fig:silhouette} shows the Silhouette coefficient results of the $k$-means clustering algorithm. Thus, we categorized crypto projects into two groups because the clustering algorithm produced the best silhouette value when $k = 4$. 

\newpage
\section{11 security common vulnerabilities and exposures}
\label{appendix:CVE}

Table~\ref{table:bitcoin_CVEs} shows Bitcoin's 11 security common vulnerabilities and exposures (CVEs) examined in the paper.

\begin{table*}[!ht]
\centering
\caption{Bitcoin's 11 security vulnerabilities examined in this work. Common Vulnerability Scoring System (CVSS) is the most widely used standard for quantifying the severity of security vulnerabilities.}
\label{table:bitcoin_CVEs}
\resizebox{0.98\linewidth}{!}{
\begin{tabular}{|l|p{0.97\linewidth}|c|}
\hline
\multicolumn{1}{|c|}{CVE-ID}         &\multicolumn{1}{|c|}{Description} &  CVSS  \\ \hline
\multirow{4}{*}{CVE-2012-1909}  & This vulnerability is a denial of service (unspendable transaction). By leveraging the ability to create a duplicate coinbase transaction, the vulnerability allows remote attackers to exploit it because the Bitcoin protocol (bitcoind, wxBitcoin, Bitcoin-Qt, and other programs) does not handle multiple transactions with the same identifier.  & \multirow{4}{*}{5.0}\\ \hline
\multirow{4}{*}{CVE-2012-1910}  & This vulnerability is a denial of service (application crash) or execution arbitrary code. Through manipulated Bitcoin protocol messages, the vulnerability in Bitcoin-Qt allows a remote attacker to exploit it because Bitcoin-Qt on Windows does not handle MinGW multithread-safe exception handling. & \multirow{4}{*}{7.5}\\ \hline
\multirow{3}{*}{CVE-2012-3789}  & This vulnerability is a denial of service (process hang). Through unknown behavior on a Bitcoin network, unspecified vulnerability in bitcoind and Bitcoin-Qt allows a remote attacker to exploit it. & \multirow{3}{*}{5.0}\\ \hline
\multirow{3}{*}{CVE-2012-2459}  & This vulnerability is a denial of service (block-processing outage and incorrect block count). Through unknown behavior on a Bitcoin network, unspecified vulnerability in bitcoind and Bitcoin-Qt allows a remote attacker to exploit it. & \multirow{3}{*}{5.0}\\ \hline
\multirow{4}{*}{CVE-2013-4165}  & This vulnerability is a timing side-channel attack. When authentication fails upon detecting the first incorrect byte of a password, the HTTPAuthorized function in bitcoinrpc.cpp in bitcoind 0.8.1 leaks information, and it allows a remote attacker to determine passwords more easily. & \multirow{4}{*}{4.3}\\ \hline
\multirow{2}{*}{CVE-2013-4627}  & A weakness in bitcoind and Bitcoin-Qt 0.8.x allows remote attackers to conduct a denial of service attack via tx message data causing excessive memory consumption. & \multirow{2}{*}{5.0}\\ \hline
\multirow{4}{*}{CVE-2014-0160}  & Both the TLS and DTLS implementations in OpenSSL 1.0.1 before 1.0.1g contain weaknesses related to the Heartbleed bug. Specially crafted packets allow remote attackers to trigger a buffer over-read and gain access to sensitive information including private keys in some instances. & \multirow{4}{*}{5.0}  \\ \hline
\multirow{5}{*}{CVE-2014-0224}  & Versions of OpenSSL before 0.9.8za, 1.0.0 before 1.0.0m, and 1.0.1 before 1.0.1h are susceptible to the ``CCS Injection'' vulnerability because they do not properly restrict ChangeCipherSpec messages. Man-in-the-middle attackers may force SSL to use a zero-length master key in some situations, allowing the attacker to gain control of SSL sessions or obtain sensitive information. & \multirow{5}{*}{6.8} \\ \hline
\multirow{6}{*}{CVE-2018-12356} & Simple Password Store 1.7.x before 1.7.2 has a vulnerability where the signature verification routine parses the output of GnuPG with an incomplete regular expression. Remote attackers may then modify the configuration file to inject additional encryption keys under their control. From this weakness, attackers may either pilfer the passwords or modify the extension scripts to conduct arbitrary code execution attacks. & \multirow{6}{*}{7.5}   \\ \hline
\multirow{2}{*}{CVE-2018-17144} & This vulnerability is a denial of service (application crash). Through duplicate input by miners, Bitcoin Core and Bitcion Knots allows remote attackers to exploit it. & \multirow{2}{*}{7.5}   \\ \hline
\multirow{5}{*}{CVE-2019-6250}  & ZeroMQ libzmq (a.k.a 0MQ) 4.2.x and 4.3.x before 4.3.1 contains a pointer overflow with code execution, which allows an authenticated attacker to overwrite an arbitrary amount of bytes beyond the bounds of a buffer. The memory layout allows the attacker to inject OS commands into a data structure located immediately after the problematic buffer, leading to arbitrary code execution attacks.& \multirow{5}{*}{9.0}\\ \hline
\end{tabular}
}
\end{table*}

\newpage
\clearpage

\section{Vulnerable and patched code fragment for CVE-2013-4627}
\label{appendix:Vulnerable and patched code fragment for CVE-2013-4627}

\begin{figure}[!h]
    \centering
    \includegraphics[width=1.0\linewidth]{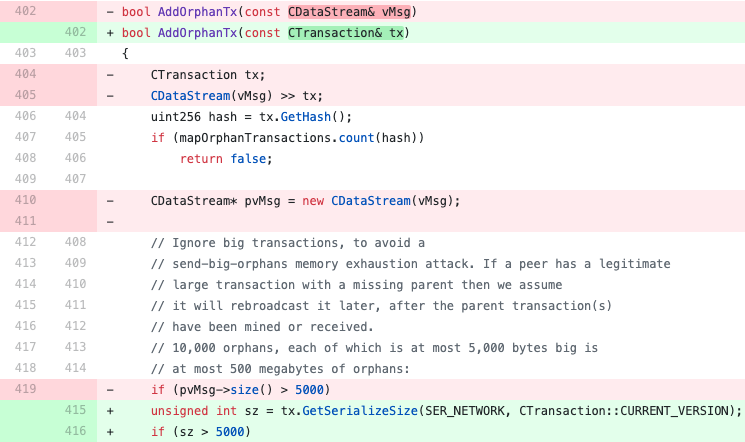}
    \caption{Vulnerable/Patched code fragment for CVE-2013-4627.}
    \label{fig:Patched code fragment}
\end{figure}

To explain what types of data were collected, we illustrate an example of vulnerable and patched code fragments for ``CVE-2013-4627,'' which is associated with a Denial of Service (DoS) attack (see Fig.~\ref{fig:Patched code fragment}). In patched code, lines starting with the ``-'' symbol have been removed, and lines starting with ``+'' symbol have been added.

When a Bitcoin node receives transaction messages, the transactions should be stored temporarily in the node's MapReplay memory even if they are invalid. Therefore, messages containing a large amount of dummy transaction data can be abused to exhaust the memory of Bitcoin nodes. To mitigate this attack, each Bitcoin node must check whether the size of the received transaction data is greater than a predefined threshold value (e.g., 5,000 bytes) to avoid large transactions. In the vulnerable code (see the lines with `-' symbol in Fig.~\ref{fig:Patched code fragment}), the size of received transaction data can directly be obtained from a field of the received message \texttt{vMsg}. Therefore, the message size checking procedure can be bypassed by assigning a small value (i.e., less than 5,000) to the message size field even when the sizes of actual transactions contained in the message can be much larger than this field value. To fix this problem, in the patched code (see the lines with ``+'' symbol in Fig.~\ref{fig:Patched code fragment}), the node compares the sizes of actual transactions (computed with \texttt{GetSerializeSize}) with the threshold value.






\section{Top 20 altcoins that are highly similar to the version of Bitcoin used for the fork}
\label{appendix:Top 20 altcoins that are highly similar to the version of Bitcoin used for the fork}

Table~\ref{table:bitcoin_children} shows a summary of the top 20 altcoins that are highly similar to the version of Bitcoin used for the fork. Interestingly, 7 (35\%) out of the 20 altcoins are contained in Cluster 4, which is frequently updated.

\begin{table*}[!ht]
\vspace{-0.5cm}
\centering
\caption{A summary of the top 20 altcoins that are highly similar to the version of Bitcoin used for the fork. ``ID'' represents the cluster ID from clustering results in Section IV. ``Market cap. (rank)'' represents each altcoin's market capitalization (and rank) in September 2021; ``--'' means that the altcoin was delisted from CoinMarketCap. ``Sim. (forked)'' represents the similarity score between the latest version of altcoin and the version of Bitcoin that was used for the fork. ``Duration'' represents the elapsed time between the time of the fork and September 3rd 2019. ``Features in white paper'' represents the key features presented in the white paper of each crypto project. ``\# commits'' represents the number of commits made after forking. ``\# vuln.'' represents the number of unpatched vulnerabilities. ``Last commit date'' represents the date of the last commit. ``R1'' indicates that there exist modifications in cryptographic algorithms and policy parameter values. ``R2'' indicates that there is no significant change. ``R3'' indicates that there is no code to implement the key features presented in the white paper of each crypto project.}
\label{table:bitcoin_children}
\resizebox{\linewidth}{!}{
\begin{tabular}{|c|c|c|c|c|r|p{3.7cm}|r|c|c|c|c|c|}
\hline
\multirow{2}{*}{Symbol} & \multirow{2}{*}{ID} & \multirow{2}{*}{Market cap. (rank)} & \multicolumn{1}{p{1cm}|}{Sim. (forked)} & \multicolumn{1}{p{1.05cm}|}{Sim. (latest)} & \multicolumn{1}{p{1.2cm}|}{Duration (days)} & \multirow{2}{*}{Features in white paper} & \multicolumn{1}{c|}{\multirow{2}{*}{\# commits}} & \multicolumn{1}{c|}{\multirow{2}{*}{\# vuln.}} & \multirow{2}{*}{Last commit date} & \multirow{2}{*}{R1} & \multirow{2}{*}{R2} & \multirow{2}{*}{R3}  \\\hline

ARY & 1 & -- & 99.4\% & 72.6\% & 612 & Supply chain management & 2 & 2 & Dec. 2017 & -- & \cmark & \cmark \\\hline

MNC & 3 & -- & 99.3\% & 64.8\% & 927 & CPU mining & 232 & 1 & Sep. 2019 & \cmark & -- & -- \\\hline

BET & 1 & -- & 99.2\% & 29.3\% & 2,140 & -- & 12 & 2 & Apr. 2014 & -- & \cmark & -- \\\hline

ACOIN & 1 & \$28k (2,509) & 98.9\% & 33.2\% & 1,828 & -- & 3 & 1 & Oct. 2014 & -- & \cmark & -- \\\hline

GAP & 1 & -- & 98.4\% & 33.2\% & 1,943 & CPU mining & 51 & 0 & Jan. 2015 & \cmark & -- & -- \\\hline

XCT & 1 & -- & 98.2\% & 57.9\% & 833 & -- & 11 & 1 & Sep. 2017 & -- & \cmark & -- \\\hline

LTC & 4 & \$15,018,401k (12) & 98.0\% & 57.9\% & 386 & CPU mining & 2,009 & 0 & Jun. 2021 & \cmark & -- & -- \\\hline

FRC & 4 & \$416k (1,827) & 97.6\% & 33.5\% & 1,943 & Tokenization & 165 & 0 & Oct. 2018 & -- & \cmark & \cmark \\\hline

PLNC & 2 & \$9k (2,602) & 97.5\% & 62.5\% & 927 & CPU mining & 191 & 2 & Jul. 2021 & \cmark & -- & -- \\\hline

CHIPS & 4 & -- & 97.5\% & 69.2\% & 754 & -- & 57 & 1 & Oct. 2018 & -- & \cmark & -- \\\hline

DEUS & 1 & -- & 97.1\% & 53.1\% & 896 & New PoW & 15 & 1 & Nov. 2017 & -- & \cmark & \cmark \\\hline

FLO & 4 & -- & 97.1\% & 70.2\% & 749 & Data storage & 192 & 2 & Feb. 2020 & \cmark & -- & -- \\\hline

MNX & 3 & -- & 97.0\% & 62.1\% & 774 & Price stability & 65 & 2 & Apr. 2019 & \cmark & -- & \cmark \\ \hline

RIC & 1 & -- & 96.9\% & 33.1\% & 2,008 & CPU mining & 402 & 1 & Jun. 2016 & \cmark & -- & -- \\ \hline

FJC & 4 & \$1,904k (1,558) & 96.9\% & 90.4\% & 185 & CPU mining & 3,817 & 0 & Aug. 2021 & \cmark & -- & -- \\ \hline

BLU & 1 & \$594k (1,942) & 96.8\% & 62.5\% & 723 & PoS & 9 & 2 & Oct. 2017 & -- & \cmark & \cmark \\ \hline

SRC & 1 & -- & 96.8\% & 28.6\% & 2,196 & Multiple hash algorithms & 2 & 4 & Aug. 2013 & -- & \cmark & \cmark \\ \hline

FTC & 4 & \$6,116k (1,155) & 96.6\% & 90.5\% & 185 & CPU mining & 188 & 0 & Jul. 2021 & \cmark & -- & -- \\ \hline

888 & 1 & -- & 96.5\% & 40.4\% & 1,726 & CPU mining & 342 & 0 & Apr. 2016 & \cmark & -- & -- \\ \hline

UFO & 4 & \$2,987k (4,746) & 96.4\% & 90.3\% & 185 & CPU mining & 153 & 0 & Mar. 2021 & \cmark & -- & -- \\ \hline
\end{tabular}
}
\end{table*}

\end{document}